\documentclass[twocolumn,showpacs,prl,amsmath,amssymb,superscriptaddress]{revtex4}

\usepackage{graphicx}
\usepackage{dcolumn}
\usepackage{bm}
\usepackage{amssymb}

\begin{document}

\title{Cooling electrons by magnetic-field tuning of Andreev reflection}

\author{Francesco Giazotto}
\email{giazotto@sns.it}
\affiliation{NEST CNR-INFM and Scuola Normale Superiore, I-56126 Pisa, Italy}

\author{Fabio Taddei}
\affiliation{NEST CNR-INFM and Scuola Normale Superiore, I-56126 Pisa, Italy}\author{Michele Governale}
\affiliation{Institut f\"ur Theoretische Physik III, Ruhr-Universit\"at Bochum, D-44780 Bochum, Germany}
\author{Carlo Castellana}
\affiliation{NEST CNR-INFM and Scuola Normale Superiore, I-56126 Pisa, Italy}
\author{Rosario Fazio}
\affiliation{International School for Advanced Studies (SISSA), via Beirut 2-4 I-34014, Trieste, Italy}
\affiliation{NEST CNR-INFM and Scuola Normale Superiore, I-56126 Pisa, Italy}
\author{Fabio Beltram}
\affiliation{NEST CNR-INFM and Scuola Normale Superiore, I-56126 Pisa, Italy}


\begin{abstract}
A solid-state cooling principle based on magnetic-field-driven tunable suppression of Andreev reflection in superconductor/two-dimensional electron gas nanostructures is proposed. This cooling mechanism can lead to very large heat fluxes per channel up to $10^4$ times greater than currently achieved with superconducting tunnel junctions. This efficacy and its availability in a two-dimensional electron system make this method of particular relevance for the implementation of quantum nanostructures operating at cryogenic temperatures.     
\end{abstract}


\maketitle

In recent times, the vast interest in mesoscopic systems brought thermoelectric phenomena
under the spotlight~\cite{giazottoRMP} in the context of solid-state refrigeration~\cite{pekolaPT}.
Promising solid-state refrigeration schemes operating at sub-Kelvin temperatures rely on superconducting microstructures~\cite{nahum} and were recently shown to yield remarkable electron~\cite{leivo} and phonon~\cite{luukanen} temperature reduction. In such systems, a normal metal (N) region is coupled to a superconductor (S) through an tunnel insulating barrier. Quasiparticle \emph{cooling} occurs because of the existence of the superconducting gap, which allows only the more energetic electrons to escape from N.
In this scheme the characteristics of the insulating barrier are crucial: on  one hand it must suppress Andreev reflection~\cite{andreev} (in order to limit NS current), on the other it must be sufficiently transmissive to maximize hot-carrier tunneling into S. Despite the remarkable performance offered by this refrigeration mechanism \cite{giazottoRMP,nahum,leivo,luukanen}, insulating barriers suffer from sub-optimal transmissivity at energies above the superconducting gap thereby limiting the efficiency of heat extraction.

Recently some of us~\cite{giazottofs} proposed to suppress  Andreev reflection while keeping highly-transmissive barriers by employing ferromagnets rather than insulators. In this Letter we present a different operational principle and investigate heat transport in superconductor/two-dimensional electron gas (2DEG) nanostructures in the presence of a localized magnetic field.
The high transmissivity available with this configuration
yields a heat current per channel that can exceed by up to a factor of $10^4$ what is currently achievable with insulating barriers. This can lead to efficient electron refrigeration, even in \emph{macroscopic} 2DEG regions. The origin  of this effect stems from the fact that a localized magnetic field applied next to a S/2DEG interface can strongly suppress Andreev reflection while affecting only marginally the above-gap transmissivity \cite{castellana}. 
Furthermore, the fine modulation of the localized field, together with the tunability of the electron-phonon interaction in 2DEGs, allow \emph{full} control of the heat current for the optimization of this refrigeration principle.

\begin{figure}[t!]
\includegraphics[width=8cm,clip]{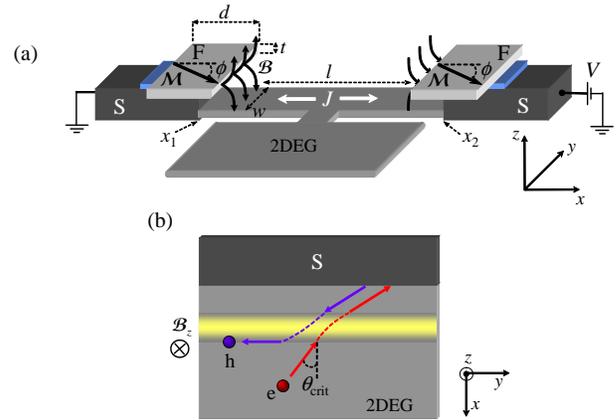}
\caption{(color) (a) Scheme of the proposed setup. Ferromagnetic strips (F) placed on top of the structure generate at each S/2DEG junction localized magnetic fields whose strength can be controlled by changing the orientation of magnetization ($\boldsymbol{\mathcal{M}}$). (b) Idealized semiclassical picture of the Andreev reflection suppression mechanism: an electron (e) impinges on the localized magnetic field (yellow region) at the critical angle ($\theta_{\text{crit}}$) above which no Andreev reflection takes place.
}
\label{fig1}
\end{figure}

The system we propose consists of a 2DEG region inserted between two identical S electrodes and is sketched in Fig. \ref{fig1}(a).  Ferromagnetic strips (F) are placed on top of each S/2DEG junction and a voltage $V$ is applied across the structure. Junctions are located at $x=x_{i}$, where $i=1,2$ corresponds to the left and right contact, respectively. $d$ long, $t$ thick F layers are positioned symmetrically on top of each junction, $z_0$ above the 2DEG [see Fig. \ref{fig1}(a)]. F layers are assumed to have homogeneous magnetization $\boldsymbol{\mathcal{M}}$. Heat and electron transport in the 2DEG region are affected by the perpendicular magnetic field produced by the $x$-component of the magnetization, $\mathcal{M}_x=|\boldsymbol{\mathcal{M}}|\text{cos}\phi$. The $z$-component of the magnetic field ($\boldsymbol{\mathcal{B}}$) can be controlled by rotating $\boldsymbol{\mathcal{M}}$ by means of a weak magnetic field \cite{johnson}. For the sake of clarity, we shall first focus on a single S/2DEG ballistic contact, namely the left one: we set $x_1=0$ and $z=0$ in the middle of the ferromagnetic layer. When $t\ll z_0$ and $t\ll d$, the $z$-component of the magnetic field generated by the left F strip in the 2DEG plane is given by \cite{matulis,governale}   
$\mathcal{B}_{z}(x)=\frac{\mu_0 t\mathcal{M}_x}{2\pi d}
[\mathcal{K}(x+\frac{d}{2},z_0)-\mathcal{K}(x-\frac{d}{2},z_0)]\theta(x)$,
where $\mathcal{K}(x,z)=-zd(x^2+z^2)^{-1}$, $\mu _0$ is the vacuum magnetic permeability, and $\theta (x)$ is the Heaviside function \cite{validity}. 
This expression for $\mathcal{B}_{z}(x)$ assumes that the magnetic field is entirely screened in the superconductor. In reality, the magnetic-field profile generated by the F strip shows a second peak located deep in the S region, which could possibly suppress superconductivity locally, but which does not affect Andreev reflection at the S/2DEG interface \cite{castellana}.
The mechanism leading to Andreev reflection suppression relies on the effect of the localized magnetic field on the orbital motion of the quasiparicles.
A simple semiclassical picture, although idealized, captures the essential features of our system. As shown in Fig.~\ref{fig1}(b), the Lorentz force acting on electrons (e) impinging on the superconductor and on Andreev-reflected holes (h) introduces a critical angle ($\theta_{\text{crit}}$) above which incoming electrons cannot be Andreev reflected.

Within the Landauer-B\"{u}ttiker scattering approach, the heat current $J(V_{\text{c}})$ through the junction biased at voltage $V_{\text{c}}$ can be written as 
\begin{eqnarray}
J=\frac{1}{h}\int^{\infty}_{-\infty} d\varepsilon \sum_{\sigma}  (\varepsilon-eV_{\text{c}})
\left\{ \left[\mathcal{N}^{\sigma}(\varepsilon)-\mathcal{R}^{\sigma}(\varepsilon)\right]
f(\varepsilon -eV_{\text{c}}) \nonumber \right. \\
\label{heatcurrent}
 \left. -\mathcal{R}_{\text{A}}^{\sigma}(\varepsilon) f(\varepsilon +eV_{\text{c}})
- \left[\mathcal{N}^{\sigma}(\varepsilon)-\mathcal{R}^{\sigma}
(\varepsilon)-\mathcal{R}_{\text{A}}^{\sigma}(\varepsilon)\right]
f(\varepsilon) \right\},
\end{eqnarray}
where $\mathcal{N}^{\sigma}$ is the number of open channels for spin-$\sigma$, $\mathcal{R}_{\text{A}}^{\sigma}$ ($\mathcal{R}^{\sigma}$) is the spin-dependent Andreev (normal) reflection probability, and $f$ is the Fermi function. 
The scattering probabilities can be evaluated numerically through a recursive Green's function technique based on a tight-binding version \cite{sanvito} of the Bogoliubov-de Gennes equation. In the absence of spin-flip scattering the latter reads
\begin{equation}
	\left( \begin{array}{cc} \mathcal{H}^{\sigma}(x) & \Delta(x) \\ \Delta^*(x) & 	-\mathcal{H}^{\sigma *}(x)
	\end{array}\right)
	\left( \begin{array}{c} u^{\sigma}\\v^{-\sigma}\end{array}\right)=\varepsilon
	\left( \begin{array}{c} u^{\sigma}\\v^{-\sigma}\end{array}\right) ,
\label{BdG}
\end{equation}
with
$\mathcal{H}^{\sigma}=(\textbf{p}-e\textbf{A}(x))\frac{1}{2m(x)}(\textbf{p}-e\textbf{A}(x))+
\mathcal{V}_{\sigma}(x)+\mathcal{U}(x)-\varepsilon_{\text{F}}^{\text{S}}$ ,
where $\textbf{p}=-i\hbar\stackrel{\rightarrow}{\nabla}$,  $\textbf{A}(x)=A_y(x)\hat{y}$ is the vector potential in the London gauge,
and $u$ ($v$) is the coherence factor for electron- (hole)-like excitations of energy $\varepsilon$, measured from the S chemical potential $\varepsilon_{\text{F}}^{\text{S}}$. The potential $\mathcal{U}(x)$ describes subband mismatch between S and the 2DEG. The Zeeman splitting in the semiconductor is given by $\mathcal{V}_{\sigma}(x)=\frac{1}{2}\sigma g^{*}\mu_{\text{B}}B(x)$, where $\mu_{\text{B}}$ is the Bohr magneton, and $g^*$ is the 2DEG effective g-factor.
The quasiparticle mass and the S pairing potential are given, respectively, by $m(x)=m^* \theta(x)+m_{\text{e}}\theta(-x)$ and $\Delta (x)=\Delta \theta(-x)$, where $m_{\text{e}}$ ($m^*$) is the free-electron (effective 2DEG) mass.
Within the tight-binding description, $\mathcal{H}^{\sigma}$ and $\Delta$ are matrices with elements $(\mathcal{H})_{ij}=\omega_i \delta_{ij}-\gamma\delta_{\left\{i,j\right\}}$ and $(\Delta)_{ij}=\Delta_0\delta_{ij}$, where $\omega_i$ is the on-site energy at site $i$, and the $\gamma$ is the hopping potential ($\left\{\cdots\right\}$ stands for first-nearest-neighbor sites) \cite{parameters1}. The presence of the magnetic field is included in the hopping potentials through Peierls phase factors \cite{Peierls}, thus taking into account the possibility of Landau level quantization in the 2DEG.

\begin{figure}[t!]
\includegraphics[width=\columnwidth,clip]{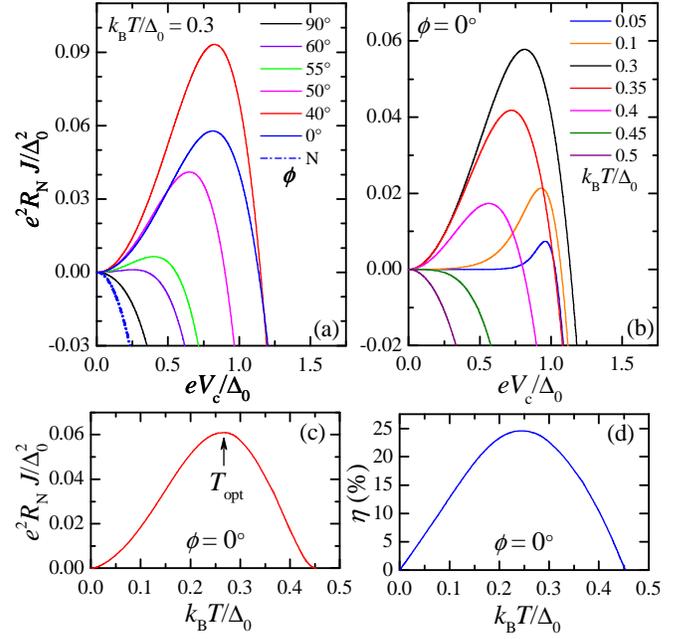}
\caption{(color) (a) Heat current $J$ of a ballistic S/2DEG system extracted from the 2DEG side vs bias voltage $V_{\text{c}}$ applied across the junction, calculated for several angles $\phi$ at $T=0.3\,\Delta_0/k_\text{B}$. $k_{\text{B}}$ is the Boltzmann constant.
Dash-dotted line represents $J$ for a N/2DEG junction at $\phi =0^{\circ}$. (b) $J$ vs $V_{\text{c}}$ calculated for several temperatures at $\phi =0^{\circ}$.  (c) $J$ calculated at the optimal bias voltage across the junction vs $T$ at $\phi =0^{\circ}$. $J$ is maximized at $T=T_{\text{opt}}\simeq 0.25\,\Delta _0/k_\text{B}$. (d) Efficiency ($\eta$) of the cooling process calculated at the optimal bias voltage vs $T$ at $\phi =0^{\circ}$. The junction normal state resistance is $R_{\text{N}}(\phi =0^{\circ})=1.07\times 10^4 \,\Omega$. }
\label{tS}
\end{figure}

Figure 2(a) displays how a strong enhancement of heat transfer off the 2DEG region occurs
by decreasing $\phi$ from $\phi=90^{\circ}$, i.e. by increasing the perpendicular magnetic field intensity. This is due
to the increasing suppression of Andreev reflection \cite{andreev} at the junction induced by the magnetic fringe field \cite{castellana} [see also Fig. \ref{fig1}(b)]. Notably, $J$ is positive (implying heat extraction from the 2DEG) at $\phi =60^{\circ}$, and is maximized around $\phi \simeq 40^{\circ}$ where it reaches the value $J\simeq 9\times 10^{-2}\,\Delta^2_0/e^2R_{\text{N}}$ corresponding to $\sim 0.5$~pW$\mu$m$^{-1}$, with $R_{\text{N}}(\phi =0^{\circ})=1.07\times 10^4 \,\Omega$ the junction normal state resistance. The smooth and gradual variation of $J$ for $0^{\circ}\leq \phi \leq 60^{\circ}$,  demonstrates that a precise  tunability of the heat flux is possible.
At each angle, $J$ is maximized by an optimal $V_{\text{c}}$. For $0^{\circ} \leq \phi \leq 40^{\circ}$ this optimal bias   is $\simeq 0.8\,\Delta_0 /e$. 
The behavior of $J$ vs $V_{\text{c}}$ for several temperatures at $\phi=0^{\circ}$ is shown in Fig. 2(b). At each temperature, $J$ is maximized by an optimal bias voltage which, at the lowest temperatures, is $V_{\text{c}}\simeq \Delta_0/e$.
Figure 2(c) shows $J$ vs temperature, calculated at each optimal bias voltage for $\phi =0^{\circ}$. In particular, $J$ is maximized at $T\equiv T_{\text{opt}}\simeq 0.25\,
\Delta_0/k_{\text{B}}$ where it reaches $J_{\text{max}}\simeq 6\times 10^{-2}\,\Delta^2_0/e^2R_{\text{N}}$. 
It is interesting to compare this S/2DEG cooling scheme with the more conventional metal/insulator/superconductor (NIS) junction configuration.
For the S/2DEG system, the junction normal-state transmissivity is $\mathcal{T}=2\pi \hbar (e^2 R_{\text{N}}\mathcal{N}_{\text{F}})^{-1}\simeq 4.4\times 10^{-2}$, where $\mathcal{N}_{\text{F}}=54$ is the total number of open channels at the Fermi energy, and the maximum heat current is $J_{\text{max}}\simeq 9\times 10^{-2}\,\Delta^2_0/e^2R_{\text{N}}$ [see Fig.~2(a)].
Considering a state-of-the-art Al$_{2}$O$_3$ tunnel junction with specific resistance $R_{\text{s}}=100\,\Omega \mu$m$^2$ \cite{leivo,luukanen},  corresponding to a normal-state transmissivity $\mathcal{T}_{\text{tunnel}}\simeq 5\times 10^{-6}$ \cite{junction}, the maximum heat flux is $(J_{\text{max}})_{\text{NIS}}\simeq 6\times 10^{-2}\,\Delta^2_0/e^2R_{\text{N}}$ at $T=0.3\,\Delta_0/k_{\text{B}}$ \cite{bardas}. Comparing the two cases, one can readily obtain that 
$\frac{J_{\text{max}}}{\mathcal{N}_{\text{F}}}\simeq \frac{3}{2} \frac{\mathcal{T}}{\mathcal{T}_{\text{tunnel}}}\left(\frac{J_{\text{max}}}
{\mathcal{N}_{\text{F}}}\right)_{\text{NIS}}\simeq 1.2\times 10^4 \left(\frac{J_{\text{max}}}
{\mathcal{N}_{\text{F}}}\right)_{\text{NIS}}$ 
i.e.,  the cooling power per channel in the S/2DEG is up to 4 orders of magnitude larger than currently attainable with insulating barriers. The difference in the heat-transport properties  originates entirely from the magnetic-field control on the S/2DEG contact which, in contrast to an insulating barrier, strongly suppresses the sub-gap conductance while only marginally affecting the junction normal-state resistance.

An important figure of merit of any refrigeration process is the  efficiency $\eta(V_{\text{c}})=J(V_{\text{c}})[V_{\text{c}}I(V_{\text{c}})]^{-1}$ \cite{giazottoRMP}, where $I$ 
is the electric current flowing through the junction. The efficiency $\eta$ versus $T$ calculated at each optimized bias voltage for $\phi=0^{\circ}$ is displayed in Fig. 2(d). Efficiency values
as high as 25\% can be obtained at $T\simeq 0.25\,\Delta_0/k_{\text{B}}$. \begin{figure}[t!]
\includegraphics[width=\columnwidth,clip]{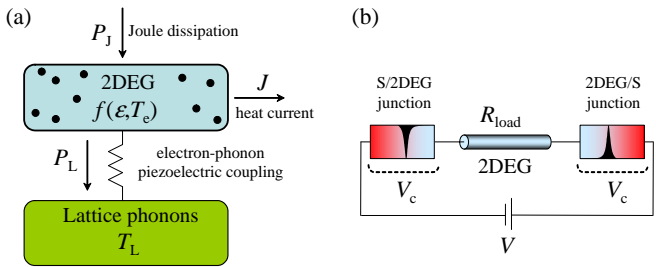}
\caption{(color) (a) Sketch of the system considered: the 2DEG is assumed to be in \emph{quasiequilibrium} at temperature $T_\text{e}$,  and to interact with the lattice phonons at temperature $T_{\text{L}}$ through \emph{piezoelectric} coupling ($P_{\text{L}}$). Thermal load due to Joule dissipation ($P_{\text{J}}$) in the 2DEG, as well as the heat flux ($J$) flowing through the system are indicated with additional arrows. (b) System equivalent electric circuit: the 2DEG is shown as a lumped resistor  ($R_{\text{load}}$), and $V$ is the total bias voltage applied across the symmetric structure.}
\label{tF}
\end{figure}

These results can be readily transferred to the case here of interest, electronic \emph{refrigeration} in S/2DEG/S structures  biased at voltage  $V$ [see Fig. 1(a)]. We focus on the \emph{quasiequilibrium} limit  \cite{giazottoRMP}, where strong electron-electron interaction thermalizes  the 2DEG at an effective electron temperature $T_{\text{e}}$ that may  differ significantly from that of the lattice  $T_{\text{L}}$ [see the scheme in Fig. 3(a)]. This limit can be attained by assuming a large enough $l$ [see Fig. \ref{fig1}(a)] so that $l_{\text{e}}<l\ll l_{\text{L}}$, with $l_{\text{e}}$ ($l_{\text{L}}$) the electron-electron (electron-phonon) scattering length \cite{lin}.
At low lattice temperatures (typically below $1~$K) \cite{BG}, \emph{piezoelectric} electron-phonon  scattering  is the main mechanism that transfers energy into III-V 2DEG systems \cite{price,ma, experiments},  and this regime will be considered in the following.
We model for simplicity the semiconductor region as a lumped element of resistance $R_{\text{load}}$ connected to the junctions [see the system equivalent circuit in Fig. 3(b)].
The stationary $T_{\text{e}}$ reached by the 2DEG is obtained by solving
\begin{equation}
\left\{
\begin{array}{l}
2J(V_{\text{c}},T_{\text{e}},T_{\text{L}})+P_{\text{L}}(T_{\text{e}},
T_{\text{L}})-P_{\text{J}}(V_{\text{c}},T_{\text{e}},T_{\text{L}})=0  \\
2V_{\text{c}}+R_{\text{load}}I(V_{\text{c}},T_{\text{e}},T_{\text{L}})-V=0,~~~~~~~~~~~~~~
\end{array}
\right.
\label{cool}
\end{equation}
where the first equation requires that the sum of all thermal fluxes in the 2DEG is zero, while the second is Kirchhoff's voltage law for the whole system [see Fig. 3(b)].
Moreover, $P_{\text{L}}(T_{\text{e}},
T_{\text{L}})$ is the energy-exchange rate due to piezoelectric coupling \cite{piezo},  $P_{\text{J}}=R_{\text{load}}I(V_{\text{c}},T_{\text{e}},T_{\text{L}})^2$ is the Joule dissipation in the 2DEG, and the factor $2$ in the first of (\ref{cool}) accounts for the fact that there are two identical junctions.
We implicitly assumed that the electron temperature of the S electrodes equals $T_{\text{L}}$ \cite{giazottoRMP}. 

In the following [see Fig. 1(a)] we choose parameters typical of In$_{0.75}$Ga$_{0.25}$As 2DEGs \cite{parameters2}. 
\begin{figure}[t!]
\includegraphics[width=\columnwidth,clip]{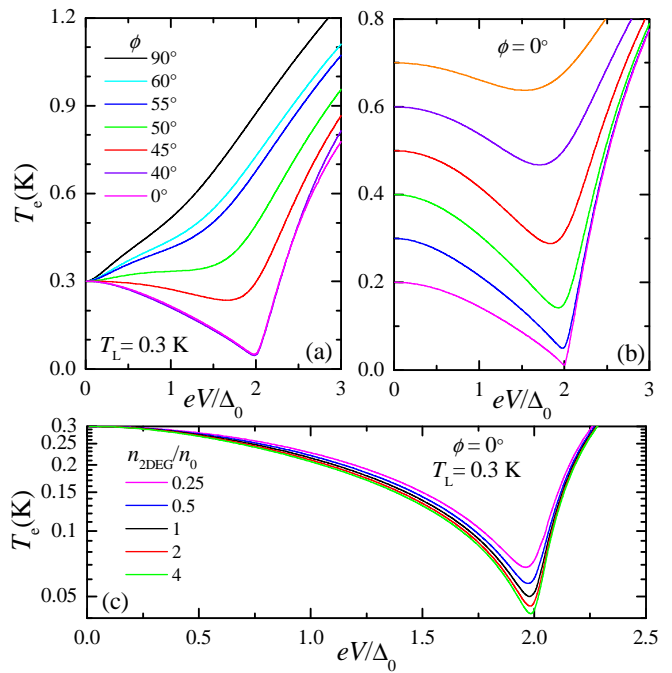}
\caption{ (color) (a) Electron temperature $T_{\text{e}}$ in the 2DEG of a S/2DEG/S refrigerator vs $V$  calculated for several angles $\phi$ at  $T_{\text{L}}=0.3$ K. (b) $T_{\text{e}}$ vs $V$ for different lattice temperatures $T_{\text{L}}$ at $\phi = 0^{\circ}$. (c) $T_{\text{e}}$  vs $V$ for several charge densities of the 2DEG ($n_{\text{2DEG}}$) at $\phi = 0^{\circ}$ and $T_{\text{L}}=0.3$ K.}
\label{exch}
\end{figure}
Figure 4(a) shows the electron temperature $T_{\text{e}}$ in the 2DEG vs  bias voltage $V$ as obtained from the solution of (\ref{cool}), for several angles $\phi$ at $T_{\text{L}}=0.3$~K. By reducing $\phi$, $T_{\text{e}}$ can be drastically lowered with respect to $T_{\text{L}}$. The refrigeration process turns out to be finely tunable for $0< \phi <50^{\circ}$ with a minimum $T_{\text{e}}\simeq 50$~mK at $V\simeq 2\Delta_0/e$ for $\phi\leq40^{\circ}$. The electronic temperature $T_{\text{e}}$ for different lattice temperatures at $\phi=0^{\circ}$ is displayed in Fig. 4(b) ($T_{\text{L}}$ values coincide with the $T_{\text{e}}$ value at $V=0$). This cooling principle makes it possible to reach a minimum $T_{\text{e}}$ around $12~$mK starting from $T_{\text{L}}=0.2~$K (temperature reduction $\sim 95\%$). Even for lattice temperatures as large as $T_{\text{L}}=0.5$~K,  $T_{\text{e}}$ reductions of $\sim 42\%$ can be achieved. 
As can be inferred from the piezoelectric energy-exchange rate $P_{\text{L}}(T_{\text{e}},
T_{\text{L}})$ (see Ref. \cite{piezo}), $\mathcal{F}_{\text{piezo}}(T)$ can be changed by varying the charge density in the 2DEG (due to the $k_{\text{F}}^{-3}$-dependence), thus  leading to a $n_0^{-1/2}$-dependence of $P_{\text{L}}$. This is a relevant peculiarity of 2DEGs as compared to normal metals. 
Figure 4(c) shows $T_{\text{e}}$ vs $V$ at $\phi =0^{\circ}$ and $T_{\text{L}}=0.3~$K calculated for several values of the charge density ($n_{\text{2DEG}}$) in the range $\simeq 1.1\times 10^{15}-1.75\times 10^{16}~$m$^{-2}$. Changes in the minimum $T_{\text{e}}$  of about $30~$mK (at $V\simeq 2\Delta_0/e$) can be obtained by  varying $n_{\text{2DEG}}$.

In summary, we have analyzed heat transport in S/2DEG nanostructures, showing that a large heat flux per channel can be achieved in the presence of localized magnetic fields.  
The precise tunability offered by  magnetic (with fields as low as some mT \cite{johnson}) as well as electrostatic control gives enhanced   freedom for the control of electronic temperatures. Materials as In$_{x}$Ga$_{1-x}$As (for $x\geq 0.75$) \cite{sorba} or InAs \cite{InAs} 2DEGs (providing Schottky barrier-free contacts with a metal) in combination with Al and Co appear as ideal candidates for applications ranging from fully-tunable refrigerators to sensitive radiation detectors \cite{giazottoRMP} operating at cryogenic temperatures. 
Partial financial support from MIUR under FIRB program RBNE01FSWY is acknowledged.


\begin{thebibliography}{99}
\bibitem{validity}
 This expression for $\mathcal{B}_z(x)$ is still a very good approximation even when $t$ and $z_0$ are of the same order \cite{governale}.
\bibitem{parameters1}
We set $\omega_i\equiv\omega^*=15.97$ eV and $\gamma\equiv \gamma^*=4.0$~eV in the 2DEG (yielding a Fermi energy $\varepsilon^*_{\text{F}} =4\gamma^*-\omega^*  =30~$meV), while $\omega_i\equiv\omega^{\text{S}} =2.4$~eV and $\gamma\equiv \gamma^{\text{S}} =1.0$~eV in S (yielding a Fermi energy $\varepsilon^{\text{S}}_{\text{F}} =1.6~$eV). This choice corresponds to a 2DEG with charge density $n_0 \simeq 4.38\times 10^{15}$~m$^{-2}$, effective mass $m^* =0.035\,m_{\text{e}}$ and $g^* =-20$, which describes well a realistic In$_{0.75}$Ga$_{0.25}$As (or InAs) 2DEG \cite{2DEG}. Moreover, we assumed a junction width $w=0.5\,\mu$m, $d=1.5\,\mu$m, $t=0.3\,\mu$m, $z_0 =0.185\,\mu$m. For the S region we chose aluminum (Al) with $\Delta_0 (T=0)=180\,\mu$eV, and  for the  F layer cobalt (Co) with a saturation magnetization $\mu _0 |\boldsymbol{\mathcal{M}}|=1.8$~T.
\bibitem{lin}
	 Note that $l_{\text{e}}$ in 2DEGs is of the order of some $\mu$m.
\bibitem{parameters2}
	In particular, we set \cite{InGaAs} $m^* =0.035\,m_{\text{e}}$, $n_0=4.38\times 10^{15}~$m$^{-2}$, $s_{\text{l}}=4.4\times 10^{3}$~ms$^{-1}$, $s_{\text{t}}=2.5\times 10^{3}$~ms$^{-1}$, $\rho =5.2\times 10^{3}$~kgm$^{-3}$, $h_{14}=1.26 \times 10^{9}$~eVm$^{-1}$, and $\epsilon ^*=14.03$. Furthermore, we chose a 2DEG region of length $l=10\,\mu$m, electron mobility $\mu =30\,$m$^2$V$^{-1}$s$^{-1}$ yielding $R_{\text{load}}=l(w n_0e\mu)^{-1}=9.48\times 10^{2}\,\Omega$, and a total area $\mathcal{A}=20\,\mu$m$^2$ [which includes a large cold-finger-like region as shown in Fig. 1(a)].
\end{thebibliography}
\end{document}